\documentclass[prl,showpacs,twocolumn]{revtex4}
\usepackage{array,amsmath,amssymb,graphicx}

\begin{document}



\title{Fluctuation and Commensurability Effect of Exciton Density Wave}

\author{Sen Yang$^1$, L.V. Butov$^1$, B.D. Simons$^2$, K.L. Campman$^3$, and A.C. Gossard$^3$}
\affiliation{$^1$Department of Physics, University of California at San Diego, La Jolla, CA 92093-0319, USA}
\affiliation{$^2$Cavendish Laboratory, Madingley Road, Cambridge CB3 OHE, United Kingdom}
\affiliation{$^3$Materials Department, University of California at Santa Barbara, Santa Barbara, CA 93106-5050, USA}

\date{\today}

\begin{abstract}

At low temperatures, indirect excitons formed at the in-plane electron-hole interface in a coupled quantum well structure undergo a spontaneous transition into a spatially modulated state. We report on the control of the instability wavelength, measurement of the dynamics of the exciton emission pattern, and observation of the fluctuation and commensurability effect of the exciton density wave. We found that fluctuations are strongly suppressed when the instability wavelength is commensurate with defect separation along the exciton density wave. The commensurability effect is also found in numerical simulations within the model describing the exciton density wave in terms of an instability due to stimulated processes.

\end{abstract}


\maketitle

An indirect exciton (IX) is a bound pair of an electron and a hole confined in spatially separated quantum well layers \cite{Lozovik1975, Fukuzawa1990}. Long lifetimes allow indirect excitons to cool down below the temperature of quantum degeneracy giving an opportunity to study low-temperature exciton states. Remarkable phenomena in cold IX gases including a spontaneous transition into a spatially modulated exciton state \cite{Butov2002, Alloing2014}, spontaneous coherence and condensation of excitons \cite{Yang2006, High2012, Alloing2014}, perfect Coulomb drag \cite{Nandi2012}, long-range spin currents and spin textures \cite{High2013, Alloing2014}, enhanced exciton radiative recombination \cite{Butov1998}, tunneling recombination \cite{Spielman2000, Eisenstein2004}, and scattering \cite{Butov2001} rates, and correlation phenomena \cite{Karmakar2009, Remeika2009, Gorbunov2012, Gorbunov2013, Schinner2013, Shilo2013, Stern2014} have been found.

\begin{figure}[h]
\includegraphics[width=0.41 \textwidth]{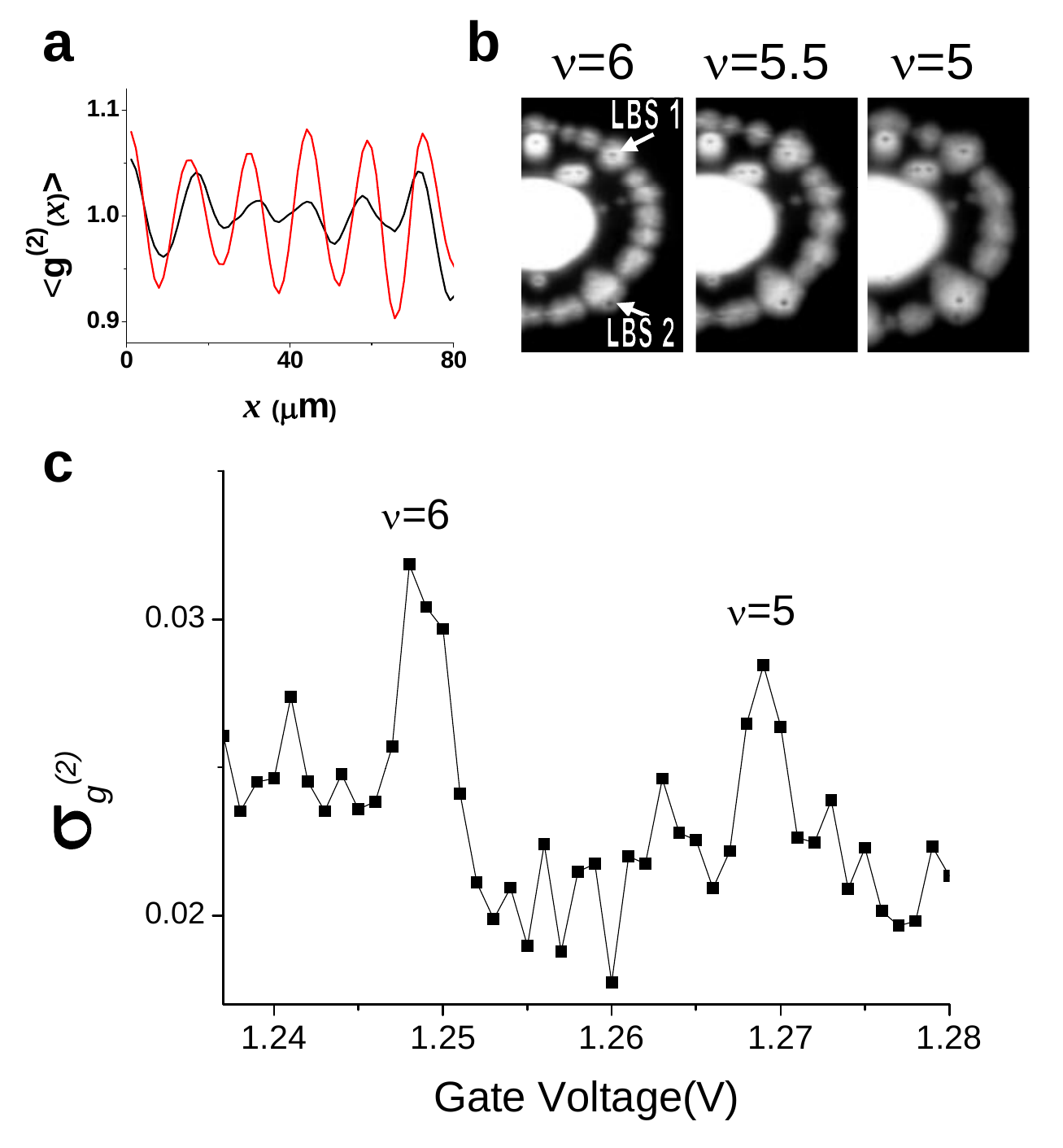}
\caption{(color online). (a) The second order correlation function for the exciton density wave $g^{(2)}(x) = \frac{\langle I(x')I(x'+x) \rangle} {\langle I(x')\rangle^2}$ for the IX emission intensity profile $I(x)$ along the ring segment between LBS 1 and LBS~2 of length $L$ [shown in (b)] with averaging over 800 frames in 27 second data acquisition movie. The commensurability numbers $\nu = L/\lambda_{\rm c}$ are 7 (red, light) and 6.5 (black, dark). (b) Images of the emission pattern of indirect excitons averaged over the 800 frames for different $\nu$. (c) Standard deviation of $g^{(2)}$ as a function of gate voltage, which controls the instability wavelength $\lambda_{\rm c}$. The peaks indicate the suppression of phase fluctuations of the exciton density wave at integer $\nu$.
} \label{1}
\end{figure}

Recently, a spatially ordered excitonic state was observed in which exciton density undergoes modulational instability \cite{Butov2002}. This state, dubbed the macroscopically ordered exciton state (MOES), exhibits approximately periodic spatial
modulation occuring within an exciton ring. The MOES forms when the IX gas is cooled below a few Kelvin close to the temperature of quantum degeneracy ($T_{\rm dB} = 2\pi\hbar^2 n/m \simeq 3$\,K for the exciton density per spin state $n = 10^{10}$ cm$^{-2}$ and exciton mass $m=0.22m_0$ relevant to the experiments). The MOES is characterized by a high degree of exciton coherence, with the coherence length reaching micrometers \cite{Yang2006, High2012, Alloing2014}. The large exciton coherence length is an order of magnitude greater than in a classical exciton gas showing that the MOES is a condensate in momentum space.

The occurrence of spatial modulation in an excitonic system \cite{Butov2002} initiated intensive experimental \cite{Yang2006, Yang2007, High2012, Alloing2014} and theoretical \cite{Yang2004, Levitov2005, Chernyuk2006, Paraskevov2007, Liu2009, Wilkes2012, Andreev2013, Andreev2014} studies. The following properties are important for understanding the MOES origin. The MOES forms in the external ring of the exciton pattern formation. The external ring itself forms on the interface between the electron-rich and hole-rich regions \cite{Butov2004, Rapaport2004, Chen2005, Haque2006, Yang2010}. The existance of such interface is essential for the MOES occurrence and, for instance, no spontaneous density modulation is observed in a cold IX gas in another ring of the exciton pattern formation -- the inner ring, which forms due to exciton transport and cooling and does not involve the border between the electron-rich and hole-rich regions \cite{Butov2002, Ivanov2006}. The other important property is that the MOES is characterized by repulsive exciton interaction \cite{Yang2007}. This is consistent with the predicted repulsive interaction between IXs, which are dipoles with a built-in dipole moment \cite{Yoshioka1990}. A search for a mechanism responsible for the formation of the MOES had led to a model attributing an instability to stimulated processes of exciton formation at the interface between the electron-rich and hole-rich regions that build up near quantum degeneracy \cite{Levitov2005}.

In this work, we report on the observation of fluctuations of the exciton density wave and finding the commensurability effect: The fluctuations vanish when the number $\nu$ of wavelengths of the exciton density wave confined between defects is an integer. This new phenomenon in cold exciton gases is presented in Figure~1. As detailed further in the text, the suppression of fluctuations of the exciton density wave at integer $\nu$ is revealed by pronounced maxima in the standard deviation of the second order correlation function for the exciton density wave $g^{(2)}(x)$ for the IX emission intensity profile $I(x)$ along the ring segment between defects. We also analyzed the stability of the exciton density wave by numerical simulations and found the commensurability effect within the model describing the exciton density wave in terms of an instability due to stimulated processes.

The coupled quantum well (CQW) structure contains two 8 nm GaAs QWs separated by a 4 nm Al$_{0.33}$Ga$_{0.67}$As and surrounded by 200 nm Al$_{0.33}$Ga$_{0.67}$As barrier layers (for details see \cite{Butov2002}). IXs in the CQW are formed from electrons and holes confined in the separated QWs. Photoexcitation was done by cw 633 nm HeNe laser with a 5~$\mu$m spot. The small disorder in the CQW is indicated by the IX emission linewidth of about 1~meV in the ring. The experiments were performed at $T=1.6$ K. IX emission images were acquired by a CCD camera after an $800 \pm 5$ nm interference filter matching the IX energy.

\begin{figure}[h]
\includegraphics[width=0.45\textwidth]{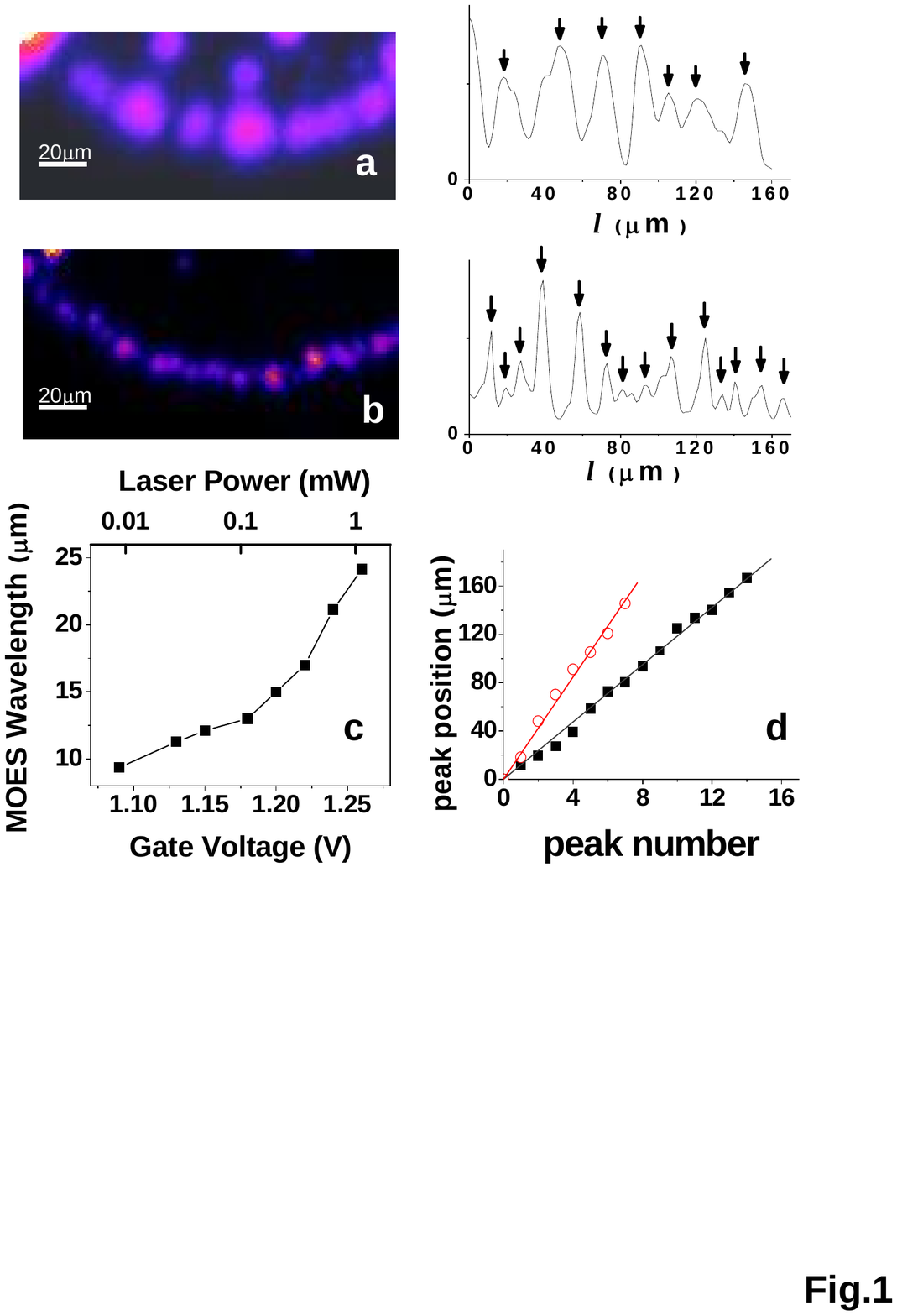}
\caption{(color online). (a,b) Left: A segment of the external ring in the exciton emission pattern. Right: The corresponding IX emission intensity profile along the ring. (a) Gate voltage $V = 1.26$ V, laser excitation power $P = 1.12$ mW. (b) $V = 1.13$ V, $P = 0.028$ mW. (c) The MOES wavelength $\lambda_{\rm c}$ vs $V$ and $P$, which are varied simultaneously so that the ring radius is fixed. (d) The MOES bead position vs peak number for the ring shown in (a) (circles) and (b) (squares).}
\label{2}
\end{figure}

Previous studies have shown that increasing the laser excitation power $P$ leads to the increase of the external ring radius due to the enhancement of hole source, while increasing the applied gate voltage $V$ leads to the decrease of the ring radius due to the enhancement of electron source \cite{Butov2004, Rapaport2004, Chen2005, Haque2006, Yang2010}. Here, we vary $P$ and $V$ simultaneously so that the ring radius and position are kept constant. This simultaneous increase of $P$ and $V$ leads to the enhancement of both electron and hole sources and, as a result, exciton density in the ring. Figure~2 shows that increasing the exciton density leads to an increase of the MOES wavelength $\lambda_{\rm c}$. Note that MOES beads are essentially equidistant forming an ordered array, while the bead intensities vary from bead to bead (Fig.~2). We refer to such quasiperiodic array as the excitonic density wave. $\lambda_{\rm c}$ is controlled by $P$ and $V$ within the range 9 -- 24~$\mu$m in the experiments presented in Fig.~2. Larger values of $\lambda_{\rm c}$ up to $\sim 40$~$\mu$m were achieved for other ring radii set by other values of $P$ and $V$.

\begin{figure}[h]
\includegraphics[width=0.3\textwidth]{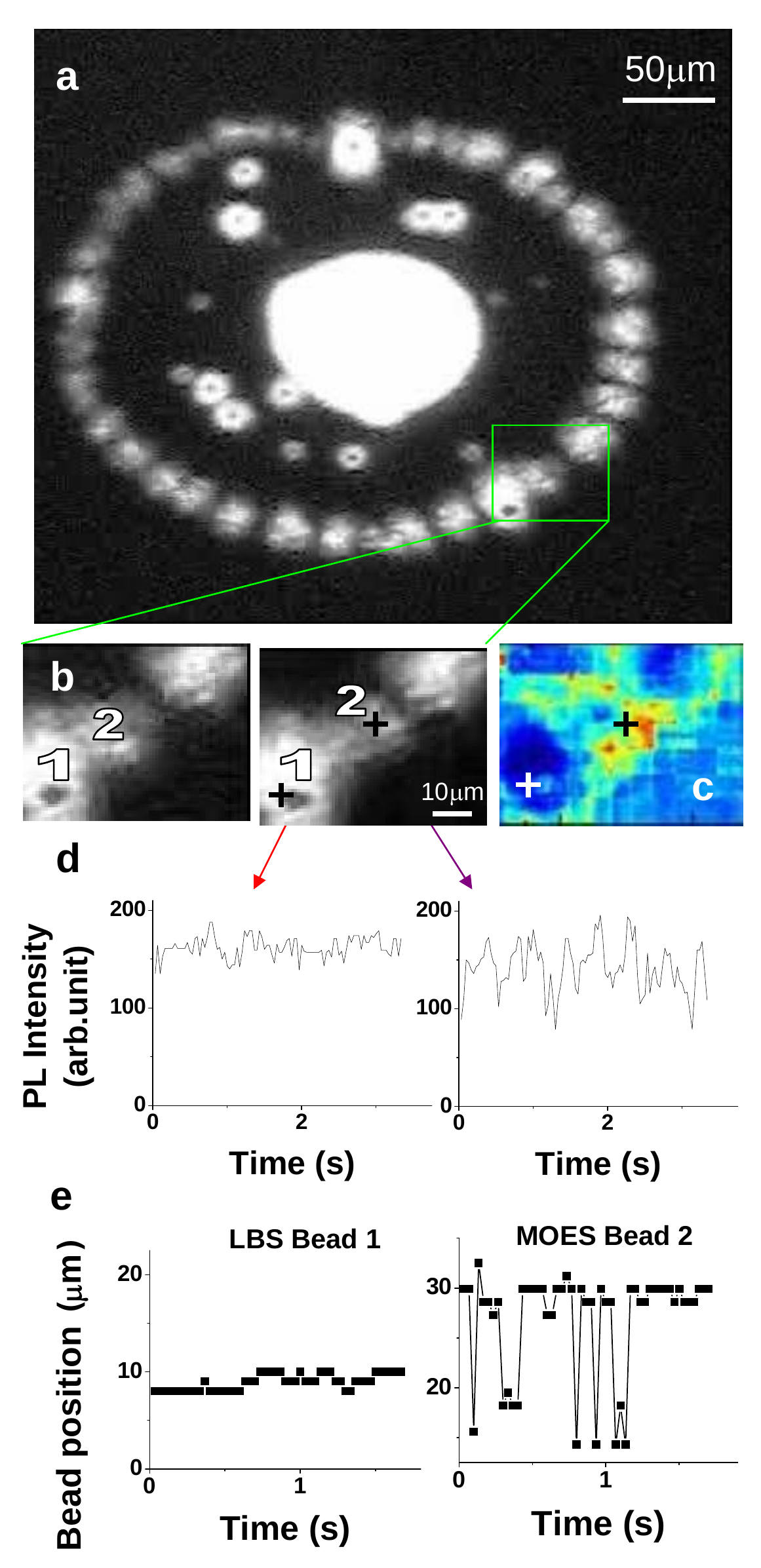}
\caption{(color online). (a) An image of the IX emission pattern extracted from a real time movie, which shows fluctuations of the exciton density wave. (b) MOES bead positions fluctuate in time: left and right images are measured at the same parameters vs time. (c) Standard deviation of the IX emission intensity. Yellow (light) color indicates high fluctuation regions and blue (dark) color indicates low fluctuation regions. (d) IX emission intensity vs time in the points marked by crosses in (b) and (c) around LBS bead (left) and MOES bead (right). (e) Bead position vs time for LBS bead (left) and MOES bead (right). $V=1.217$ V. MOES beads fluctuate while LBS beads are stable.} \label{3}
\end{figure}

The exciton pattern formation also includes localized bright spots (LBS), which are associated with defects in the sample -- electron current filaments \cite{Butov2004}. LBS beads are clearly distinct from MOES beads: the positions of LBS beads are fixed while MOES beads move with the ring when the position of the laser spot is adjusted, besides LBS beads have hot cores associated with the current-induced heating while MOES beads don't \cite{Butov2004}. Figure~3 and a movie in supplementary materials show that LBS beads are stable while MOES beads fluctuate with time. Both the observed fluctuations of the exciton density wave (Fig.~3) and its wavelength variation with density for the fixed ring position (Fig.~2) indicate that the exciton density modulation in the MOES forms spontaneously rather than due to the in-plane disorder in CQW.

\begin{figure}[h]
\includegraphics[width=0.4\textwidth]{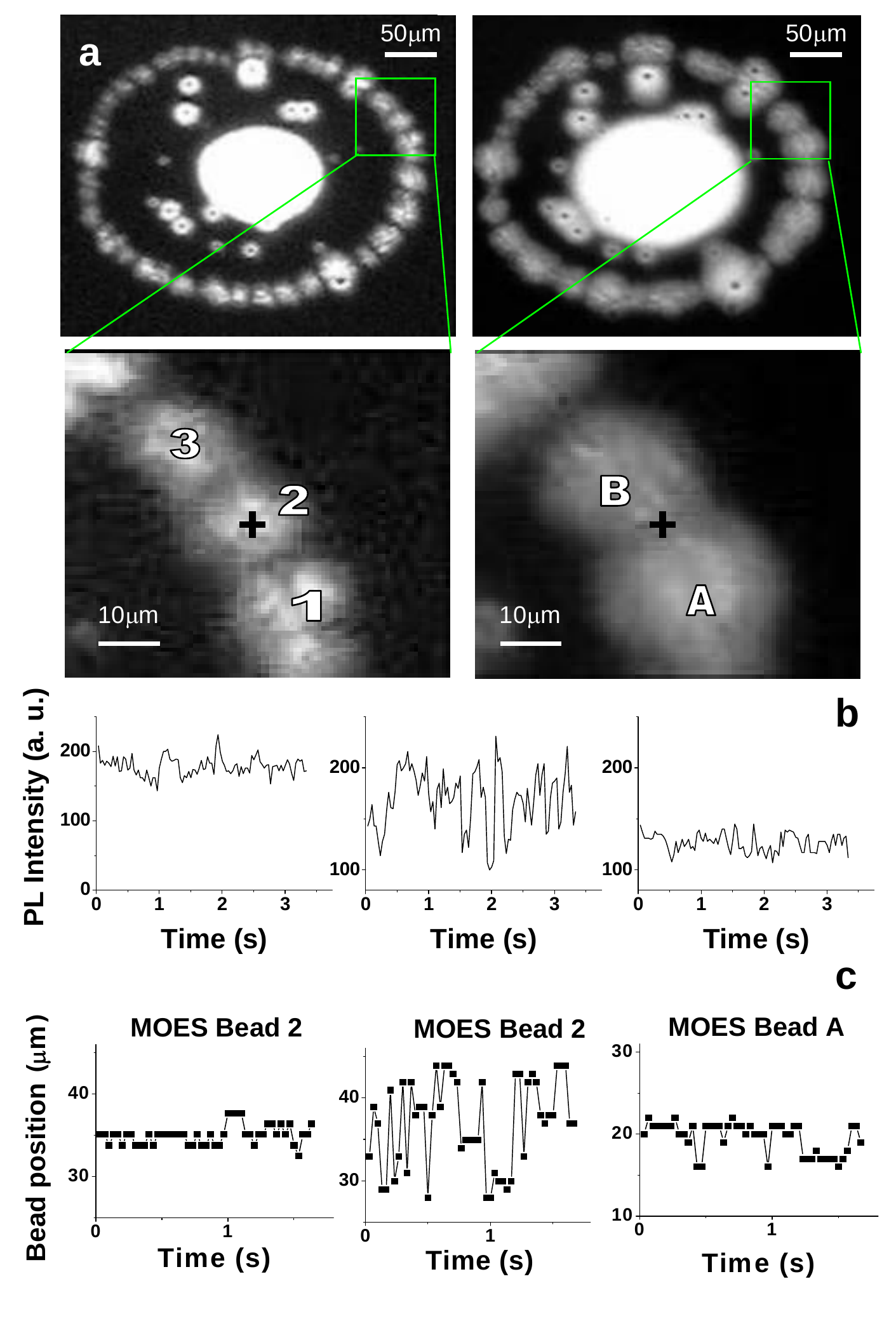}
\caption{(color online). (a) Images of the IX emission pattern for different MOES wavelengths $\lambda_{\rm c}$. The commensurability number $\nu = L/\lambda_{\rm c}$ in the ring segment between LBS 1 and LBS 2 of length $L$ is $\nu = 8$ (left) and 6 (right). (b) Fluctuations of the IX emission intensity for integer (left and right) and non-integer (middle) $\nu$ in the point marked by a cross in (a). (c) Fluctuations of the bead positions for the same conditions as in (b). $V = 1.219$ (left), $1.233$ (middle), and $1.263$ (right) V. Fluctuations of MOES beads vanish at integer $\nu$.} \label{4}
\end{figure}

The stability of LBS beads and fluctuations of MOES beads (Fig.~3) show that the phase of the exciton density wave in the ring is locked at LBS defects and fluctuates in between them. Controlling the exciton density in the ring (by varying $P$ and $V$) allows to probe the fluctuations of the exciton density wave for different ratios between the MOES wavelength $\lambda_{\rm c}$ and the length $L$ of the ring segment between two LBS on the ring (such as LBS 1 and 2 in Figs.~1 and 4). Figure~4 shows that the amplitude of the fluctuations is small when the number of wavelengths of the exciton density wave confined between the defects $\nu = L/\lambda_{\rm c}$ is an integer. In turn, fluctuations increase for non integer $\nu$, compare in Fig.~4b,c the medium panel presenting large fluctuations at non integer $\nu$ with left and right panels presenting smaller fluctuations at integer $\nu$.

This commensurability effect is quantified in Fig.~1, which presents the second order correlation function for the exciton density wave $g^{(2)}(x) = \frac{\langle I(x')I(x'+x) \rangle} {\langle I(x')\rangle^2}$ for the IX emission intensity profile $I(x)$ along the ring segment between LBS~1 and 2. Apparently, a stable periodic wave produces strong oscillations in $g^{(2)}(x)$, while fluctuations of the wave smear out such oscillations. Standard deviation of $g^{(2)}$ gives a measure for the fluctuations. Figure~1 shows pronounced maxima in $g^{(2)}$ indicating suppression of the fluctuations of the exciton density wave at integer $\nu$.

The observed fluctuations of the exciton density wave and commensurability effect are discussed below. The MOES is a state with spontaneously broken symmetry. It involves a large number of excitons, for instance the estimated number of excitons in the ring segment between LBS~1 and LBS~2 is $\sim 10^6$. The commensurability effect indicates that the fluctuations of the exciton density wave are collective. Collective fluctuations in states with spontaneously broken symmetry is a general phenomenon observed in a variety of systems. Rotational fluctuations and waves in liquid crystals, sound waves in liquids and solids, and second sound waves in superfluids present characteristic examples. Here, we compare the experimental results for the exciton density wave with the theory attributing the development of the MOES to stimulated processes that build up near quantum degeneracy \cite{Levitov2005} and show that the experimentally observed commensurability effect is also found within this model.

This instability is encapsulated by a simple kinetic theory involving the interplay of the electron/hole and exciton densities. In particular, the dimensionless electron density, $g_{\rm e} \equiv \frac{D_{\rm e}}{c\ell}n_{\rm e}$, satisfies the nonlinear diffusion equation,
\begin{eqnarray*}
\nabla^2 g_{\rm e}=\exp\left[\eta
\delta g_{\rm x}\right]g_{\rm e}(g_{\rm e}-x)\,,
\end{eqnarray*}
where lengths are measured in terms of the diffusion length, and fluctuation of the local exciton density from its value in the unmodulated steady state, $\delta g_{\rm x}\equiv g_{\rm x}-\bar{g}_{\rm x}$, depends non-locally on the fluctuation in electron density $\delta g_{\rm e}\equiv g_{\rm e}-\bar{g}_{\rm e}$ through the relation, $\delta g_{\rm x}(\vec{x})=-\delta g_{\rm e}(\vec{x}) +\int \frac{d^2x'}{2\pi\ell_{\rm x}^2} K_0(\frac{|\vec{x}-\vec{x}'|}{\ell_{\rm x}})\delta g_{\rm e}(\vec{x}')$, with $K_0$ the modified Bessel function. Here $c$ denotes the total carrier flux at the interface, $D_{\rm e}$ denotes the electron diffusion constant, and the control parameter, $\eta$, involves both the proximity to the degeneracy temperature and exciton density of the unmodulated state at the electron-hole interface (for details, see \cite{Levitov2005}).

In the ring geometry, an analysis of the kinetic equation above shows that, below a critical temperature, the ring undergoes a type II instability toward the development of a spatially modulated state with a wavelength set by the length scale $\lambda_{\rm c}\sim \ell^{1/3}\ell_{\rm x}^{2/3}$, but constrained to be commensurate with the overall ring circumference (in the relevant parameter regime, the diffusion length $\ell_{\rm x}$ exceeds the range of electron and hole overlap $\ell$). To assess the potential for collective fluctuations to drive the observed commensurability effect, we explored the pattern of instability for a fixed value of $\eta=10$, where the instability is well-developed, for changing values of $\ell/\ell_{\rm x}$. When the instability is allowed to ``anneal'' from the unmodulated state, the wavenumber, $\lambda_{\rm c}$, of the modulated state changes sequentially through a sequence of plateaus set by discrete values compatible with the ring circumference (Fig.~5). However, when the system is allowed to evolve from the fully-modulated state as a function of changing $\ell/\ell_{\rm x}$, one sees both hysteresis and the exclusion of an intermediate stable wavenumber. These effects mirror the commensurability effect seen in experiment -- the stochastic transfer between stable and metastable states a feature of discontinuous transitions, and the exclusion of stable modulations -- a manifestation of the nonlinearity of the dynamics.

The experiment shows also that both the MOES wavelength $\lambda_{\rm c}$ and ring width $\delta_{\rm r}$ increase with density (Fig.~2). This data is also consistent with the model where both $\lambda_{\rm c}$ and $\delta_{\rm r}$ increase with density [$\lambda_{\rm c}\sim \ell^{1/3}\ell_{\rm x}^{2/3}$ and $\delta_{\rm r} \sim \l_{\rm x}$ and $\ell_{\rm x}$ and $\ell$ increase with density due to screening of the in-plane disorder].

\begin{figure}[h]
\includegraphics[width=0.4\textwidth]{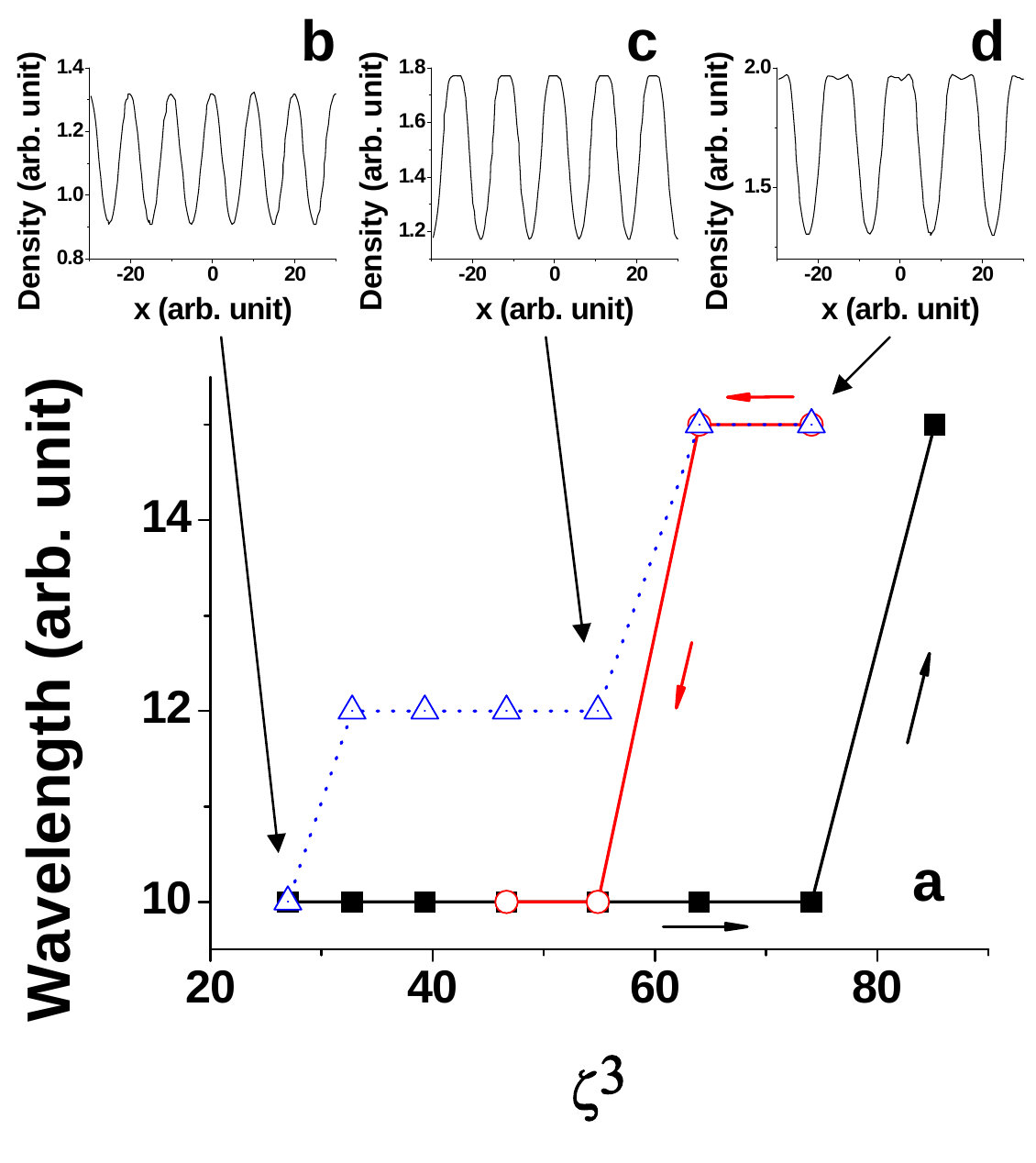}
\caption{(color online). (a) Wavelength of the exciton density wave as a function of $(\ell_{\rm x}/\ell)^3$. Squares (circles) show the evolution of $\lambda_{\rm c}$ as $\ell_{\rm x}/\ell$ is ramped up progressively from 3 to 4.4 (down progressively from 4.2 to 3.6) using the previous value as the seeding density. Triangles show evolution of $\lambda_{\rm c}$ for steady-state exciton density wave as $\ell_{\rm x}/\ell$ is changed from 3 to 4.2 using a small random perturbation from the uniform solution as a seeding density. (b-d) Corresponding profiles of the exciton density wave along the electron-hole interface. Commensurate states with integer $\nu$ are found to be robust with respect to the parameter variation producing the plateaus, while fluctuations develop in the transition region between integer $\nu$ where hysteresis is found.} \label{5}
\end{figure}

In conclusion, we observed fluctuations of the exciton density wave and the commensurability effect -- the fluctuation suppression when the number of wavelengths confined between defects is an integer.

We thank Leonid Levitov for valuable discussions and contributions at the earlier stage of the exciton pattern formation studies. This work was supported by NSF.

\end{document}